\documentclass[preprint,aps,showpacs,superscriptaddress,nofootinbib]{revtex4}
\usepackage{graphicx,subfig}
\usepackage{dcolumn}
\usepackage{amsfonts, amssymb,amsmath}
\usepackage{hhline}
\usepackage{multirow}

\begin{document}

\title{Three-body calculation of the $1s$ level shift in kaonic deuterium with realistic $\bar{K}N$ potentials.}

\author{J. R\'{e}vai}
\affiliation{Wigner Research Center for Physics, RMI,
H-1525 Budapest, P.O.B. 49, Hungary}

\date{\today}

\begin{abstract}
The $1s$ level shift in kaonic deuterium was calculated using Coulomb Sturmian expansion of Faddeev equations. The convergence of the method yields
an $\sim\  1\  eV$ accuracy for the level shifts. We used three different, realistic, multichannel $\bar{K}N$ interactions reproducing all known experimental
two-body $K^-N$ data. The different results suggest, that the level shift should be in the range  $\Delta E\sim(800\pm30)-(480\pm20)i\ \ eV$.
The  exact  level shifts  were compared with values, given by  the commonly used approximations.
\end{abstract}

\pacs{13.75.Jz, 11.80.Gw, 36.10.Gv}
\maketitle

\section{Introduction}
\label{Introduction_sect}
Hadronic atoms are valuable sources of information about interaction of different, negatively charged  hadrons with nuclei and -- indirectly -- with individual
nucleons. A large amount of work, both theoretical and experimental, has been devoted to this subject. A comprehensive review of the field is presented
in the book of A. Deloff \cite{Del}, one of the starting points  of which reads:

{\it
``.. the conventional picture of hadronic atoms (is) based on a two-body model Hamiltonian in which all strong interaction effects have been
simulated by an absorptive potential representing the complicated interaction between the hadron and the nucleus...''  }

Apart from the simplest case of hadronic hydrogen, this is obviously an approximation, the validity of which to our knowledge has not been investigated yet.
The simplest case, where this can be at least attempted is the three-body system of hadronic deuterium. This particular system is also challenging from the
strangeness nuclear physics side: it can provide additional information about the basic $\bar{K}N$ interaction, unobtainable from the two-body data.
 Powerful methods exist for practically exact
solution of the three-body problem, in particular, for finding real or complex eigenvalues: Faddeev integral equations or coordinate space variational
methods. However, for the case of hadronic deuterium both have to face  serious difficulties: the Faddeev equations encounter the everlasting problem
of Coulomb interaction (especially attractive), while for the variational calculations the problem lies in the presence of two very different  -- and relevant --
distance scales.

Some years ago Z. Papp proposed a method \cite{papp1} for simultaneous treatment of short-range and Coulomb-forces in three-body systems. The
method is based on the discretization of Faddeev equations on Coulomb Sturmian (CS) basis. The method was successfully applied to short range plus
repulsive Coulomb interaction (nuclear case) and purely Coulomb systems with attraction and repulsion \cite{papp2}. The present case of three strongly interacting
hadrons with Coulomb attraction between certain pairs, which is practically inaccessible for other methods, was not considered previously.

In a short paper \cite{test} we reported the results of a test calculation to demonstrate the applicability of this method for the case of
kaonic deuterium. For simplicity, the calculations were performed with simple complex one-channel  $\bar{K}N$ potentials, the effect coupling to the
$\pi\Sigma$ channel was imitated by an energy independent absorptive part. In the present work  realistic multichannel  $\bar{K}N$  interactions
were used, which reproduce all known experimental data. In section II.  a somewhat more detailed description of the method is given with due
emphasis on the important issues of its application for multichannel systems. In sect. III we present our results, while sect. IV. contains the conclusions.

\section{Method}
\label{Method.sect}
\subsection{The basic equations}
The simplest hadronic atom in which the deviation from the conventional two-body picture can be studied is hadronic deuterium, in
our case kaonic deuterium. It is a three-body problem, for which we shall use the notations of Fig. 1.

The Hamiltonian reads:
\begin{equation*}
H=H_0+v_1^s(x_1)+v_2^s(x_2)+v_3^s(x_3)-{e^2\over x_3}P
\end {equation*}
with
\begin{equation*}
H_0=-{1\over 2\mu_i}\Delta_{x_i}-{1\over 2\mu_{i,jk}}\Delta_{y_i}=h_0(x_i)+h_0(y_i)=...
\end{equation*}
Here the $(x_i,y_i), i=1,2,3$ are the usual Jacobi coordinates, the $v_i^s(x_i)$ denote the strong interaction between the particle pairs. The indices $i$ stand for the usual Faddev partitions: the
spectator particle $i$ and the corresponding interacting pair $(jk)$. The peculiarity
of the system is, that particles in pair 3 can be in two particle states, and, accordingly,  $v_3(x_3)$ is a $2\times 2$ matrix, while $P$ is a projection operator on the
$K^-p$ particle state:
\begin{equation}
\label{v3}
v_3(x_3)=\left(\begin{array}{cc}v_{pK^-}&v_{pK^-,nK^0}\\v_{nK^0,pK^-}&v_{nK^0}\end{array}\right),\qquad
P=\left(\begin{array}{cc}1&0\\0&0\end{array}\right).
\end{equation}

\begin{figure*}
\begin{center}
\includegraphics[width=.7\textwidth]{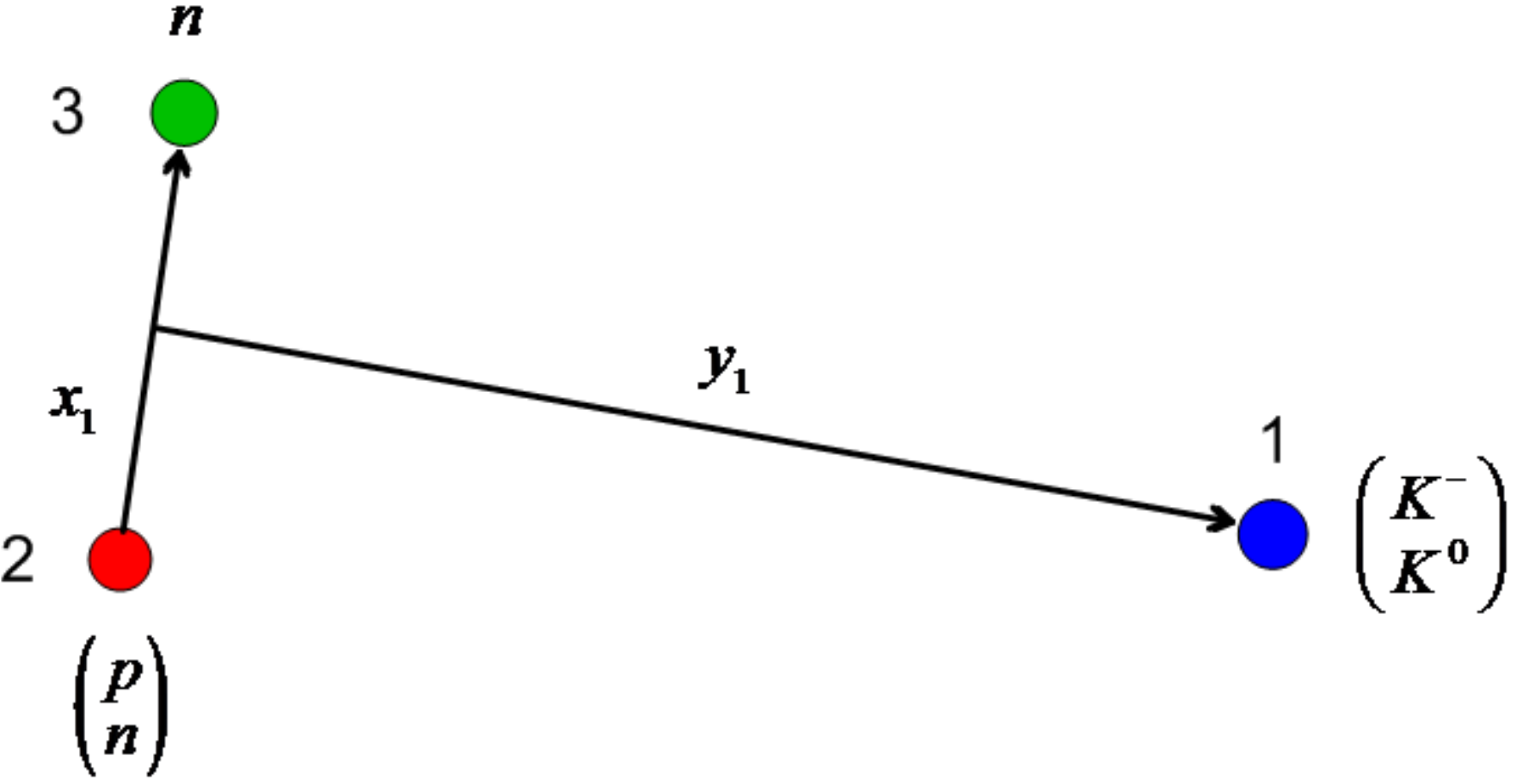}
\caption{The $K^-d$ three-body system}
\label{fig1}
\end{center}
\end{figure*}

Assuming at the first stage, that particles 1,2 and 3 are distinguishable, we have 3 coupled particle channels: $(\bar{K}^0 n_1n_2),(K^-n_1p_2)$
and $(K^-,p_1n_2)$, and correspondingly, a column wave function $\Psi$, which is then separated into the usual Faddeev components:
\begin{equation*}
\Psi=\left(\begin{array}{c}\Psi^{K^0n_1n_2}\\ \Psi^{K^-n_1p_2}\\ \Psi^{K^-p_1n_2}\end{array}\right)=
\left(\begin{array}{c}\Psi^{K^0n_1n_2}_1\\ \Psi^{K^-n_1p_2}_1\\ \Psi^{K^-p_1n_2}_1\end{array}\right)+
\left(\begin{array}{c}\Psi^{K^0n_1n_2}_2\\ \Psi^{K^-n_1p_2}_2\\ \Psi^{K^-p_1n_2}_2\end{array}\right)+
\left(\begin{array}{c}\Psi^{K^0n_1n_2}_3\\ \Psi^{K^-n_1p_2}_3\\ \Psi^{K^-p_1n_2}_3\end{array}\right).
\end{equation*}

 Coupled Faddeev equations for the 9 unknown functions can be written down, however,
symmetrization with respect to baryon indices 1 and 2 simplifies the system: symmetric and antisymmetric combinations are decoupled. Since
the deuteron is antisymmetric in these indices (the two-body isospin $I=0$), we have to work with the antisymmetric combinations. In this case the component
$\Psi^{K^0n_1n_2}_1$ disappears from the equations and from the remaining 8 functions
 4 antisymmetric combinations are left as unknown functions.
They satisfy the Noble \cite{Noble} form of homogeneous Faddeev equations, when the Coulomb interaction is added to $H_0$:
\begin{equation}
\label{psnp}
\Psi_{np}(x_1,y_1)=G_{np}(x_1,y_1;E)v_{np}(x_1)(\Psi_{K^-n}(x_2,y_2)+\Psi_{K^-p}(x_3,y_3))
\end{equation}
\begin{equation}
\label{psKn}
\Psi_{K^-n}(x_2,y_2)=G_{K^-n}(x_2,y_2;E)v_{K^-n}(x_2)(\Psi_{np}(x_1,y_1)+\Psi_{K^-p}(x_3,y_3))
\end{equation}
\begin{equation}
\label{psKp}
\left(\begin{array}{c}\Psi_{K^-p}(x_3,y_3)\\ \Psi_{K^0n}(x_3,y_3)\end{array}\right)=G_3(x_3,y_3;E)v_3(x_3)
\left(\begin{array}{c}\Psi_{np}(x_1,y_1)+\Psi_{K^-n}(x_2,y_2)\\-\Psi_{K^0n}(x_2,y_2)\end{array}\right)
\end{equation}
with
\begin{equation}
\label{gnp}
G_{np}(x_1,y_1)=\left(E-h_0(x_1)-h_0(y_1)-v_{np}(x_1)+{e^2\over |\frac{1}{2}x_1+y_1|}\right)^{-1}
\end{equation}
\begin{equation}
\label{gKn}
G_{K^-n}(x_2,y_2)=\left(E-h_0(x_2)-h_0(y_2)-v_{K^-n}(x_2)+{e^2\over |\frac{m_N}{m_N+m_K}x_2+y_2|}\right)^{-1}.
\end{equation}
 $G_3(x_3,y_3;E)$ and $v_3(x_3)$ are $2\times 2$ matrices:
\begin{equation}
\label{gKp}
G_3(x_3,y_3;E)=\left(\left(\begin{array}{cc}E-h_0(x_3)-h_0(y_3)+{e^2\over x_3}&0\\
0&E-h_0(x_3)-h_0(y_3)\end{array}\right)-v_3(x_3)\right)^{-1},
\end{equation}
while $v_3(x_3)$ was defined in eq.(\ref{v3}).  The functions and operators in eqs.(\ref{psnp})-(\ref{gKp}) are labeled by the interacting pair. It has to be noted, that
the Coulomb potential is the same in all three Green operators, expressed in different Jacobi coordinates.

 In our earlier Faddeev calculations of the $\bar{K}NN$ system \cite{preva}-\cite{prevc} we used "isospin" representation for labeling the interacting pairs, since the strong interactions are
 assumed to be isospin conserving, acting "separately" in the $I=0$ and $I=1$ two-body isospin states. In the present case, however,
 due to the presence of the Coulomb force, acting between a certain (charged) particle pair $(K^-p)$, it is preferable to work in particle
 representation. Accordingly, the two-body interactions, that occur in eqs.(\ref{psnp})-(\ref{gKp}),  have to be transformed from the $I=0$ and $I=1$ representation:
\begin{align*}
v_{nn}&=v_{NN}^{I=1};&\quad &v_{np,np}=v_{pn,pn}=(v_{NN}^{I=0}+v_{NN}^{I=1})/2;&\quad &v_{np,pn}=(v_{NN}^{I=1}-v_{NN}^{I=0})/2;\\
v_{K^-n}&=v_{\bar{K}N}^{I=1};&\quad &v_{K^-p}=v_{K^0n}=(v_{\bar{K}N}^{I=1}+v_{\bar{K}N}^{I=0})/2;&\quad &v_{K^-p,K^0n}=(v_{\bar{K}N}^{I=1}-v_{\bar{K}N}^{I=0})/2.
\end{align*}
The potentials $v_{np,pn}$ and $v_{K^-p,K^0n}$ correspond to interactions changing the identity of particles. For the $v_{np}$  the symmetrization procedure yields
\begin{equation*}
v_{np}=(v_{np,np}-v_{np,pn})=v_{NN}^{I=0},
\end{equation*}
the $I=0\ NN$ interaction, responsible for the deuteron.
\subsection{Exact optical potential}

In our previous test calculation \cite{test}, where we investigated the applicability of the method \cite{papp1} for calculating the level shift, for the interactions
occurring in eqs.(\ref{psnp})-(\ref{gKp}) we used simple absorptive one (particle) channel potentials. On the other hand, realistic calculations for the $(\bar{K}NN)$ system
require the inclusion of the strong coupling between the $\bar{K}N-\pi\Sigma$ (or even  $\bar{K}N-\pi\Sigma-\pi\Lambda$ ) channels. Therefore
in our earlier Faddeev calculations  for the $\bar{K}NN$ system without the Coulomb-interaction \cite{preva}-\cite{prevc}
\footnote{After submission of the present paper a comprehensive and detailed review of our work on the $\bar{K}NN$ system appeared \cite{Shev}.}
we explicitly treated the coupled
$\bar{K}NN-\pi\Sigma N$ particle channels. We also checked, under what conditions the coupled particle channel problem can be reduced
to the single $\bar{K}NN$ channel. We found, that replacing the multichannel $\bar{K}N$ interaction by the so called "exact optical"
potential (deduced from it), a single $\bar{K}NN$ channel Faddeev calculation yields for the observables connected with this channel (e.g.
$\bar{K}pp$ quasi-bound state or $\bar{K}d$ low-energy scattering data) results practically coinciding with those of a complete, coupled
channel calculation. Since the $1s$ level shift is also of this type, for its calculation we used the same procedure.

The ``exact optical'' potential for a given channel of a multichannel interaction is defined as a potential, exactly reproducing the diagonal
t-matrix element of the multichannel interaction in that channel. For separable interactions its construction is straightforward: it amounts to adding  an energy-dependent part to the coupling constant of the retained channel ($\bar{K}N$ in our case).
For a two-channel   $\bar{K}N-\pi\Sigma$ potential of the form
\begin{equation*}
\hat{V}=\left( \begin{array}{cc}
|g_{\bar{K}N}\rangle\lambda_{\bar{K}N}\langle g_{\bar{K}N}|\quad &|g_{\bar{K}N}\rangle\lambda_{\bar{K}N,\pi\Sigma}\langle g_{\pi\Sigma}|\\
|g_{\pi\Sigma}\rangle\lambda_{\pi\Sigma,\bar{K}N}\langle g_{\bar{K}N}|\quad &|g_{\pi\Sigma}\rangle\lambda_{\pi\Sigma}\langle g_{\pi\Sigma}|\\
\end{array}
\right)
\end{equation*}
the $\hat{V}^{opt}_{\bar{K}N}$ is
\begin{equation*}
\hat{V}^{opt}_{\bar{K}N}=|g_{\bar{K}N}\rangle\lambda^{opt}_{\bar{K}N}(E)\langle g_{\bar{K}N}|
\end{equation*}
with
\begin{equation*}
\lambda^{opt}_{\bar{K}N}(E)=\lambda_{\bar{K}N}+{\lambda_{\bar{K}N,\pi\Sigma}^2\langle g_{\pi\Sigma}|G^0_{\pi\Sigma}(E)|g_{\pi\Sigma}\rangle
\over {1-\lambda_{\pi\Sigma}\langle g_{\pi\Sigma}|G^0_{\pi\Sigma}(E)|g_{\pi\Sigma}\rangle}}
\end{equation*}
where $G^0_{\pi\Sigma}(E)$ is the free Green-operator in the excluded channel. For the 3-channel case the procedure is somewhat more complicated,
but also straightforward. The physical quantities (t-matrices, Green-operators) calculated using these exact optical potentials carry the full
analytical structure - poles, branch points and cuts - of the original multichannel interaction.
\subsection{The Coulomb Sturmian basis}
The CS functions are defined as
\begin{equation*}
\langle{\bf r}|nlm\rangle=\langle{\bf r}|\mu\rangle=N_{nl}r^le^{-br}L_n^{2l+1}(2br)Y_{lm}(\Omega_r),
\end{equation*}
where $L_n^{2l+1}$ is an associated Laguerre polynomial and $b$ is a range parameter. They are orthogonal with respect to the weight function
$1/r$, or defining their adjoint functions as $\langle{\bf r}|\tilde{\mu}\rangle=\langle{\bf r}|\mu\rangle/r$, they form a biorthogonal set
with them:
\begin{equation*}
\langle\mu|{1\over r}|\mu'\rangle=\delta_{\mu\mu'};\qquad \langle\mu|\tilde{\mu'}\rangle= \langle\tilde{\mu}|\mu'\rangle=\delta_{\mu\mu'}.
\end{equation*}
The CS basis is discrete and complete:
\begin{equation*}
\sum_{\mu=0}^{\infty}|\mu\rangle\langle\tilde{\mu}|=\sum_{\mu=0}^{\infty}|\tilde{\mu}\rangle\langle\mu|=\hat{\bf I}\approx
\sum_{\mu=0}^{N_{max}}|\mu\rangle\langle\tilde{\mu}|.
\end{equation*}
The most remarkable feature of the CS basis is, that in this representation the matrix of the operator $(z-h_c)$, where $h_c$ is the
two-body Coulomb Hamiltonian
\begin{equation*}
h_c=-{1\over 2m}\Delta_{{\bf r}}\pm {e^2\over r}
\end{equation*}
is tridiagonal:
\begin{equation*}
\langle\mu|z-h_c|\mu'\rangle={1\over 2b}\delta_{ll'}
\left\{
\begin{array}{l}
-\delta_{n,n'+1}[\sqrt{n(n+2l+1)}(z+b^2/2m)]\\
+\delta_{n,n'}[2(n+l+1)(z-b^2/2m)\mp 2be^2]\\
-\delta_{n,n'-1}[\sqrt{(n+1)(n+2l+2)}(z+b^2/2m)]
\end{array}
\right\}.
\end{equation*}
This feature allows to set up an infinite tridiagonal set of equations for the matrix elements of the Coulomb Green operator
$g_c(z)=(z-h_c)^{-1}$:
\begin{equation*}
\langle\mu|(z-h_c)g_c(z)|\tilde{\mu}'\rangle=\delta_{\mu\mu'}=\sum_{\nu=0}^{\infty}\langle\mu|(z-h_c)|\nu\rangle
\langle\tilde{\nu}|g_c(z)|\tilde{\mu}'\rangle,
\end{equation*}
which can be solved exactly \cite{papp3},\cite{Del1}. The same holds for the matrix elements of the free Green operator $g_0(z)$.

Introducing a double CS basis for each set of Jacobi coordinates:
\begin{equation*}
\langle x_iy_i|\mu\rangle_i=\langle x_i|\mu_x\rangle \langle y_i|\mu_y\rangle;\qquad \mu=(\mu_x,\mu_y)
\end{equation*}
the unknown functions $\Psi_i$ for $i=np, K^-n,K^-p,K^0n$ can be expanded on this basis:
\begin{equation}
\label{exp}
\Psi_i(x_i,y_i)=\sum_{\mu}^{N_i}\langle x_iy_i|\mu\rangle_i\  X^i_{\mu},
\end{equation}
where $X^i_{\mu}={} _i\langle\tilde{\mu}|\Psi_i(x_i,y_i)\rangle$.

\subsection{The matrix equation}
Before setting up the matrix equations for the new unknowns $X^i_{\mu}$ two intermediate steps are needed.

When operators, expressed in one set of Jacobi coordinates, act on functions depending
on another set, what is typical for Faddeev equations, we have to introduce a transformation matrix:
\begin {equation*}
\hat{O}(x_i,y_i)\Psi_j(x_j,y_j)\Rightarrow\sum_{\mu',\mu''}{}_i\langle \mu|\hat{O}(x_i,y_i)|\mu'\rangle_i\  M^{(ij)}_{\mu,\mu''}X^j_{\mu''},
\end{equation*}
where $M^{(ij)}$ is the overlap matrix of the two CS basis sets, depending on different Jacobi coordinates:
\begin{equation*}
M^{(ij)}_{\mu,\mu'}={}_i\langle\tilde{\mu}|\mu'\rangle_j,
\end{equation*}
which is energy independent and can be calculated by numerical integration.

When calculating the matrix elements of Green operators entering the eqs.(\ref{gnp})-(\ref{gKp}) two cases have to be distinguished.
In $G_3(x_3,y_3)$ the Coulomb interaction depends on its "native" relative coordinate $x_3$, thus it corresponds
to a Green operator of non-interacting two-body subsystems sharing a common 3-body energy. For this case a calculation
scheme exists. The Green operators of eqs.(\ref{gnp}),(\ref{gKn}), on the other hand, are genuine 3-body operators, due
to the Coulomb interaction, which depends on both "native" Jacobi coordinates. To make them calculable we have to
split the Coulomb interaction into "channel" and "polarization" parts:
\begin{equation*}
\frac{e^2}{|\gamma_i x_i+y_i|}=V^{ch}(y_i)+U_i(x_i,y_i);\qquad \gamma_{np}=-\frac{1}{2};\quad \gamma_{K^-n}=\frac{m_N}{m_N+m_K}
\end{equation*}
with
\begin{equation*}
V^{ch}(y_i)=\frac{e^2}{y_i};\quad U_i(x_i,y_i)=\frac{e^2}{|\gamma_i x_i+y_i|}-\frac{e^2}{y_i}
\end{equation*}
The channel potential $V^{ch}(y_i)$ is the Coulomb interaction of the spectator particle with the center of mass of the
interacting pair, while the polarization potential $U_i(x_i,y_i)$ causes distortion of the internal motion of
the pair due to the displacement of the Coulomb interaction from the charged particle to the center of mass.
For the Green operators  $G_{np}$ and $G_{K^-n}$ of eqs.(\ref{gnp}),(\ref{gKn}) the following resolvent equations
can be written down:
\begin{equation*}
G_{np}=G^{ch}_{np}+G^{ch}_{np}U_{np}G_{np}\quad {\rm and}\quad G_{K^-n}=G^{ch}_{K^-n}+G^{ch}_{K^-n}U_{K^-n}G_{K^-n},
\end{equation*}
where the channel Green operators $G^{ch}_{np}$ and $G^{ch}_{K^-n}$ were introduced:
\begin{equation}
\label{gchnp}
G^{ch}_{np}(x_1,y_1)=\left( E-h_0(x_1)-h_0(y_1)-v_{np}(x_1)+{e^2 \over y_1}\right)^{-1}
\end{equation}
\begin{equation}
\label{gchKn}
G^{ch}_{K^-n}(x_2,y_2)=\left( E-h_0(x_2)-h_0(y_2)-v_{K^-n}(x_2)+{e^2 \over y_2}\right)^{-1}.
\end{equation}

With their help the first two Faddeev equations  (\ref{psnp}),(\ref{psKn}) can be rewritten as
\begin{equation}
\label{mf1}
\begin{array}{l}
\Psi_{np}(x_1,y_1)=\\G^{ch}_{np}(x_1,y_1;E)[U_{np}(x_1,y_1)\Psi_{np}(x_1,y_1)+v_{np}(x_1)(\Psi_{K^-n}(x_2,y_2)+\Psi_{K^-p}(x_3,y_3))]
\end{array}
\end{equation}
\begin{equation}
\label{mf2}
\begin{array}{l}
\Psi_{K^-n}(x_2,y_2)=\\G^{ch}_{K^-n}(x_2,y_2;E)[U_{K^-n}(x_2,y_2)\Psi_{K^-n}(x_2,y_2)+v_{K^-n}(x_2)(\Psi_{np}(x_1,y_1)+\Psi_{K^-p}(x_3,y_3))]
\end{array}
\end{equation}

Applying now the expansion (\ref{exp}) to the modified Faddeev equations (\ref{mf1}),(\ref{mf2}) and ({\ref{psKp}) we get a matrix equation of the
 form ${\bf X}={\bf A}(E){\bf X}$ with
\begin{equation*}
{\bf X}=\left(
\begin{array}{c}
{\bf X}_{np}\\{\bf X}_{K^-n}\\{\bf X}_{K^-p}\\{\bf X}_{K^0n}
\end{array}
\right)
\end{equation*}

\begin{equation}
\label{fin}
{\bf A}(E)=\left(
\begin{array}{cccc}
{\bf G}^{ch}_{np}{\bf U}_{np}&{\bf G}^{ch}_{np}{\bf v}_{np}{\bf M}^{(12)}&{\bf G}^{ch}_{np}{\bf v}_{np}{\bf M}^{(13)}&0\\
{\bf G}^{ch}_{K^-n}{\bf v}_{K^-n}{\bf M}^{(21)}&{\bf G}^{ch}_{K^-n}{\bf U}_{K^-n}&{\bf G}^{ch}_{K^-n}{\bf v}_{K^-n}{\bf M}^{(23)}&0\\
({\bf G}_3{\bf v}_3)_{11}{\bf M}^{(31)}&({\bf G}_3{\bf v}_3)_{11}{\bf M}^{(32)}&0&-({\bf G}_3{\bf v}_3)_{12}{\bf M}^{(32)}\\
({\bf G}_3{\bf v}_3)_{21}{\bf M}^{(31)}&({\bf G}_3{\bf v}_3)_{21}{\bf M}^{(32)}&0&-({\bf G}_3{\bf v}_3)_{22}{\bf M}^{(32)}
\end{array}
\right)
\end{equation}
Here bold face letters stand for vectors and matrices in the corresponding CS basis. Our task is to find the (complex) solution of the equation
${\rm Det}(\hat{\bf I}-{\bf A}(E))=0$ close to the unperturbed value
\begin{equation}
\label{E0}
E_0=E_d+\varepsilon_{1s}(Kd),
\end{equation}
where $E_d$ is the deuteron binding energy, while $\varepsilon_{1s}(Kd)$ is the ground state energy of the $K^-$ in the Coulomb field
of a point-like deuteron. $E_0$ is the lowest bound state pole of the channel Green operator $G_{np}^{ch}$.

\subsection{Calculation of Green operator matrix elements.}
All Green operators of our final equations (\ref{fin}) are now of the form
\begin{equation*}
G(x,y;E)=(E-h_0(x)-h_0(y)-u_1(x)-u_2(y))^{-1}
\end{equation*}
with
\begin{equation*}
\begin{array}{c}
G_{np}^{ch}\Rightarrow\ u_1(x)=v_{np}(x);\quad u_2(y)=-e^2/y\\
G_{K^-n}^{ch}\Rightarrow\ u_1(x)=v_{K^-n}(x);\quad u_2(y)=-e^2/y\\
G_3\Rightarrow u_1(x)=v_3(x)-\left(\begin{array}{cc}e^2/x&0\\0&0\end{array}\right)\ \ {\rm (a\  matrix);}\quad u_2(y)=0.
\end{array}
\end{equation*}
For them the following convolution integral representation exists:
\begin{equation}
\label{conv}
G(x,y;E)=\oint_c g_1(x;\varepsilon)g_2(y;E-\varepsilon)\ d\varepsilon
\end{equation}
with
\begin{equation*}
g_1(x;z)=(z-h_0(x)-u_1(x))^{-1}\quad{\rm and}\quad g_2(y;z)=(z-h_0(y)-u_2(y))^{-1}.
\end{equation*}
In the original formulation \cite{Bianchi} the contour $c$ "encircles the spectrum of $g_1$  without
penetrating the spectrum of  $g_2$". For practical purposes this can be reformulated as ``the contour $c$ is a directed path, which divides
the complex plane into two non-intersecting parts, the singularities of $g_1$ being on its left side, while those of $g_2$ on
its right side''. Obviously, the double CS matrix elements of $\langle \mu|G|\mu'\rangle$ can be expressed in the same way through the
matrix elements of $\langle\mu_x|g_1|\mu_x'\rangle$ and $\langle\mu_y|g_2|\mu_y'\rangle$, each in its own basis.

The choice of the integration path $c$ can depend on the analytical properties of the two-body Green operators $g_1$ and $g_2$ entering the
convolution integral (\ref{conv}) and on the position of the 3-body energy $E$ on the complex plane with respect to the singularities
of  $g_1$ and $g_2$. For ``ordinary'' 3-body problems -- real energy,
bound or scattering states -- it can be chosen on the physical sheets of both $g_1$ and $g_2$.  When searching for quasi-bound states in a
3-body problem with simple absorptive potentials -- complex energy eigenvalue -- the path still can remain solely on the physical sheets.
This was the case in our previous calculation \cite{test}. In the case of looking for resonance poles in scattering -- complex eigenvalues on the
closest non-physical sheet of at least one of the $g_i$-s -- the contour has to be continued to that non-physical sheet (see e.g. \cite{papp4}).
And, finally, our present problem: quasi-bound state in a multichannel system, where one of the $g_i$-s is defined on a multilevel Riemann surface.
Before discussing this case in some detail, a technical point may be mentioned:  since the integration along the real $\varepsilon$ axis leads to strong
oscillations, especially for the high $n,n'$ matrix elements, it is desirable to keep the path as far from the real axis as the fixed branch points allow.

In Fig. 2. we show the integration path (dashed line) for the case of $G^{ch}_{np}$ of eq.(\ref{gchnp}):
\begin{equation*}
G^{ch}_{np}(x_1,y_1;E)=\oint_c g_{np}(x_1;\varepsilon)g_c(y_1;E-\varepsilon)\ d\varepsilon\ .
\end{equation*}
Here both operators $g_{np}$ and $g_c$ have only the usual unitary cuts, starting at zero energy, and one (or more) bound state poles for negative real energies (denoted by crosses).
For a 3-body energy $E=E_0$, where $E_0$ was defined in eq.(\ref{E0}), the small crosses on Fig.2., corresponding to the deuteron energy $E_d$ and the Coulomb ground state energy $\varepsilon_{1s}$ coincide, and $G^{ch}_{np}$ has a pole, as mentioned before.
Along the integration path the values of  $g_{np}$ and $g_c$ are taken  from their usual physical sheets ($Im(k)>0$).

\begin{figure}
\includegraphics[width=.7\textwidth]{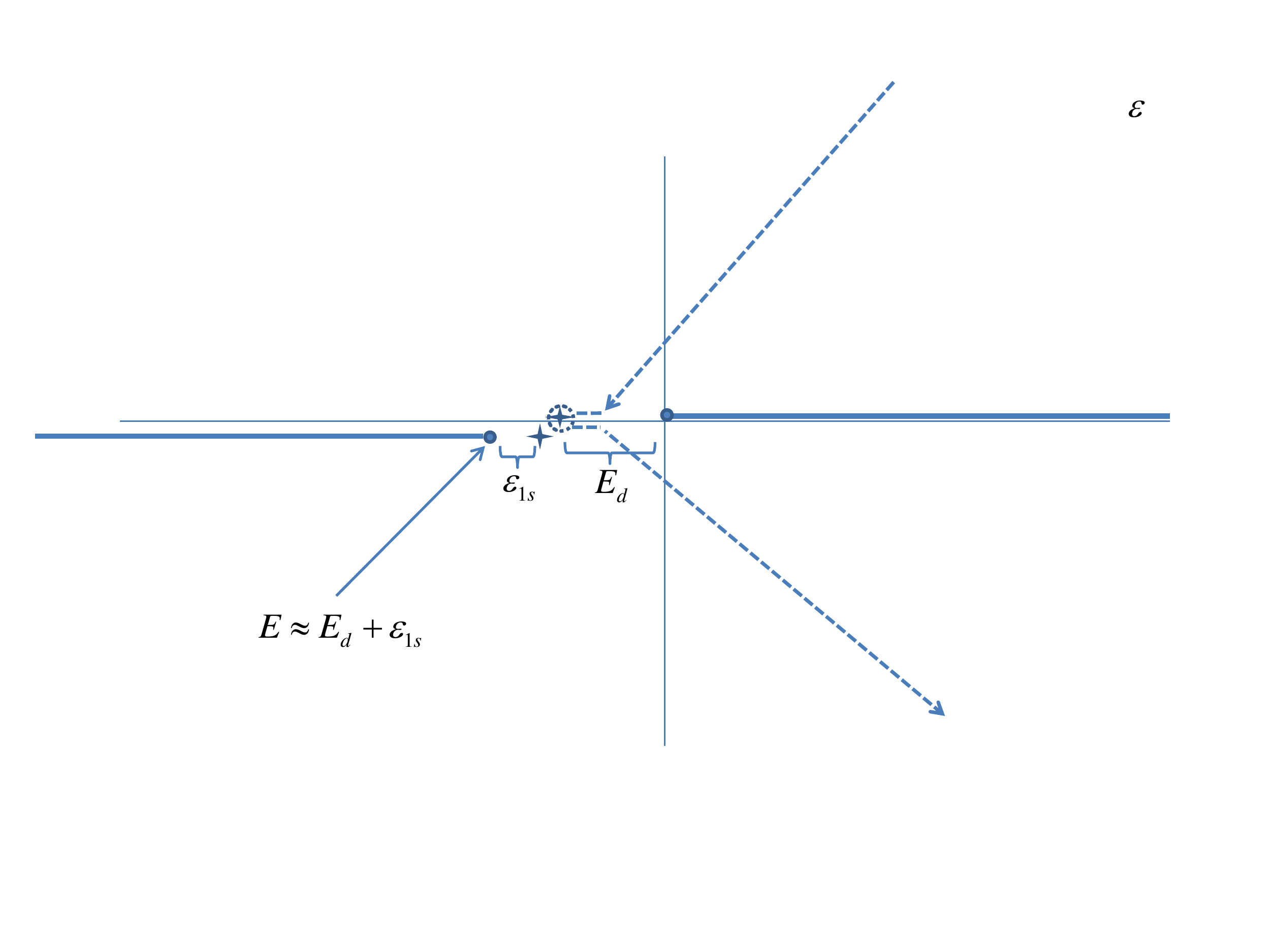}
\caption{Integration path for the channel Green operator $G^{ch}_{np}$. See details in the text.}
\end{figure}

A more complicated situation is shown on Fig. 3., the case of $G^{ch}_{K^-n}$ of eq.(\ref{gchKn}):
\begin{equation*}
G^{ch}_{K^-n}(x_2,y_2;E)=\oint_c g_{K^-n}(x_2;\varepsilon)g_c(y_2;E-\varepsilon)\ d\varepsilon\ .
\end{equation*}
Here, apart from the unitary cuts, the $g_{K^-n}$, due to the construction of the ``exact optical'' potential, ``remembers'' the corresponding cut of the excluded $\pi\Sigma$ channel, starting at the $\pi\Sigma$ threshold. With respect to this cut the sought eigenvalue $E$ is on the unphysical sheet, below the physical one. The situation with the conventional  cuts and the Coulomb pole is depicted  in Fig. 3a. Choosing the integration path in accordance with this picture, one could not avoid the undesired integration between the two cuts along the real axis (or very close to it). Therefore, with a certain redefinition of how the square root is taken in the $\pi\Sigma$ channel, the  $\pi\Sigma$ cut can be turned ``upwards'' (as shown in Fig. 3b.),
allowing to select the integration path denoted by the dashed line. Integrating along this path, the values of  $g_c$ must be taken from its physical sheet, while those of  $g_{K^-n}$ -- from its  (redefined)  unphysical sheet.   A consequence  of this latter is the possible occurrence of poles of  $g_{K^-n}$ on the ``wrong'' side of the path - their contribution has to be taken into account when evaluating the convolution integral (indicated by small dashed circles around them in Fig. 3b.)

\begin{figure}
\begin{tabular}{lc}
\includegraphics[width=.7\textwidth]{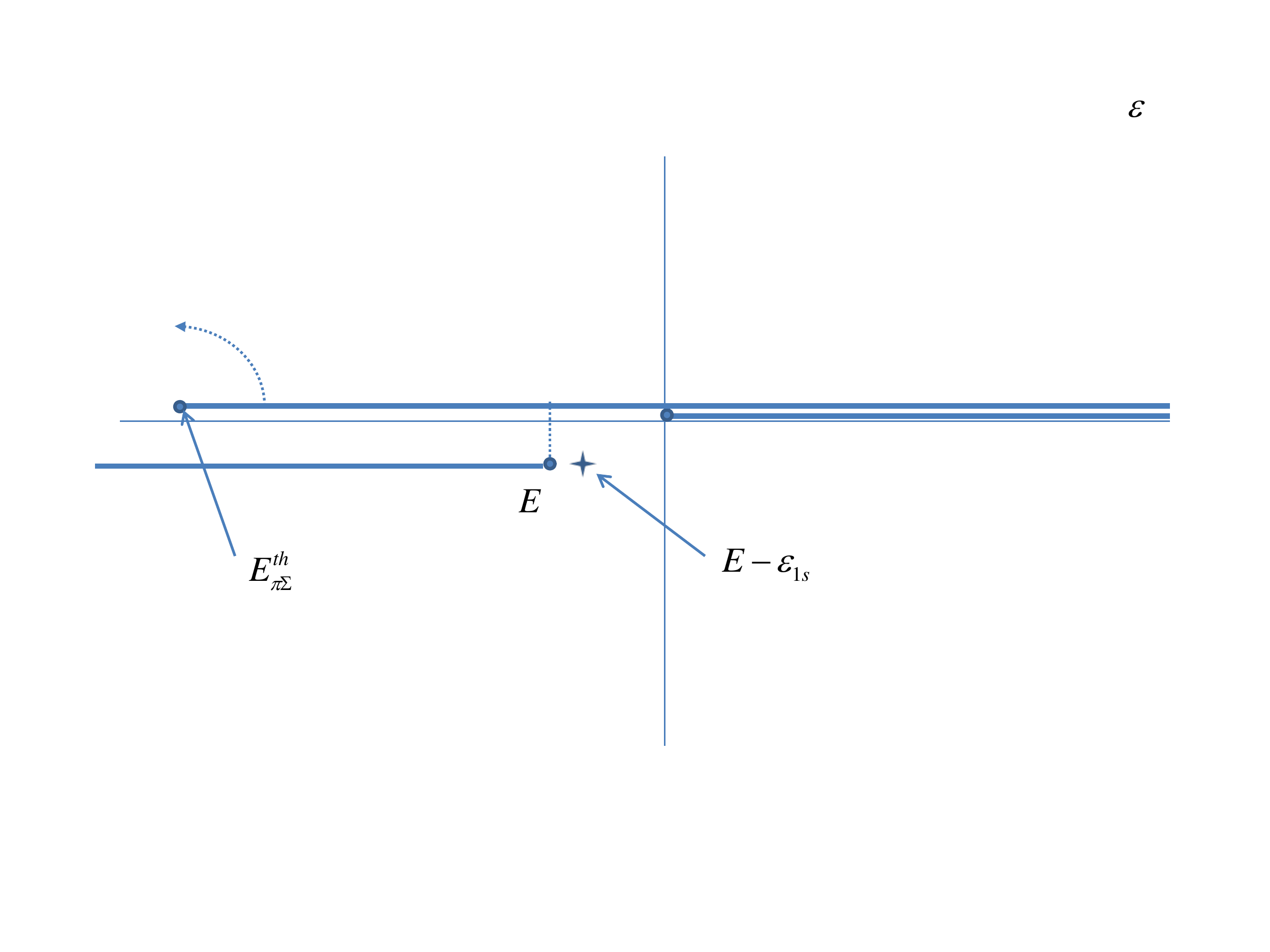}&\raisebox{7cm}{(a)}\\
\includegraphics[width=.7\textwidth]{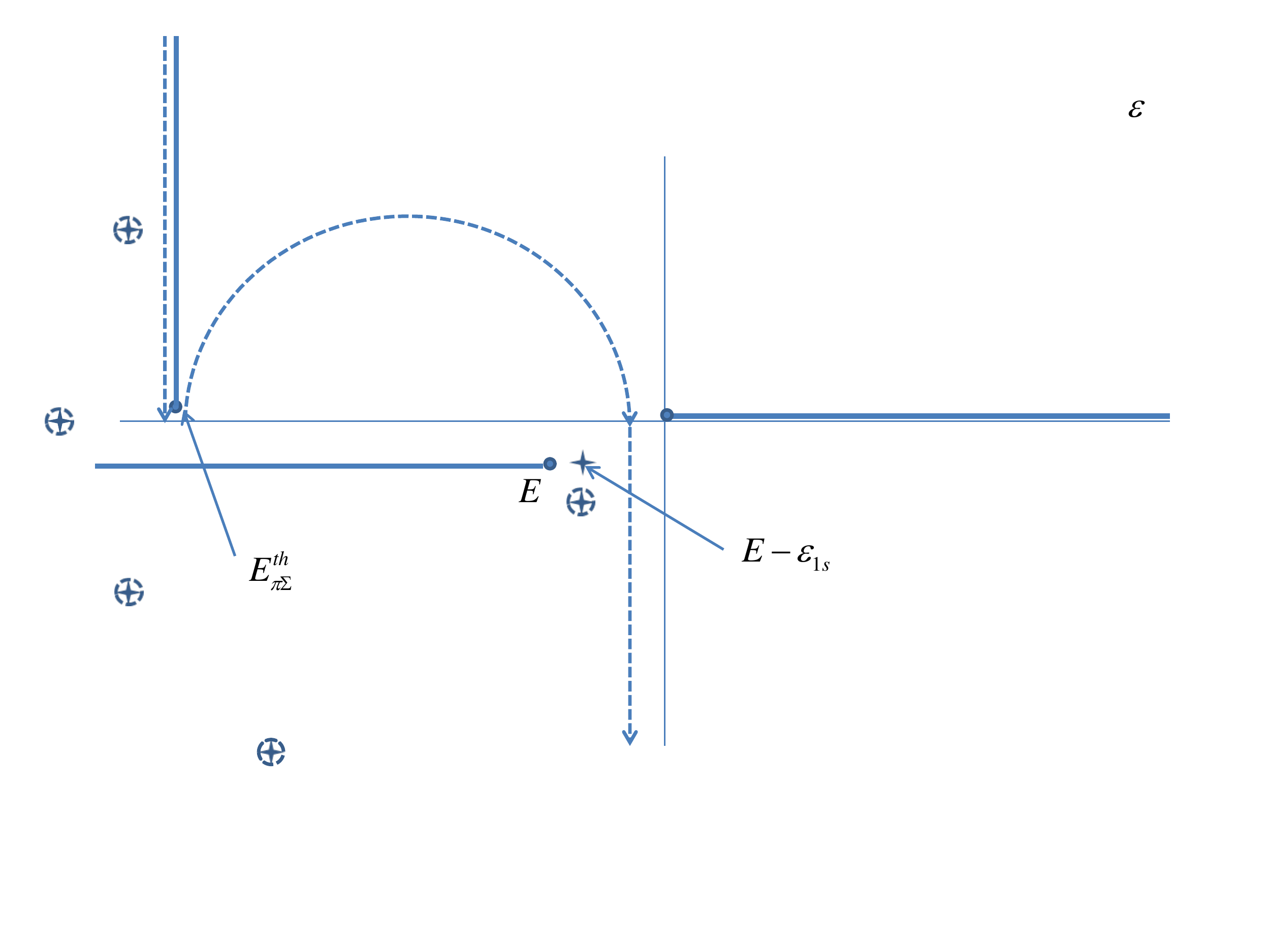}&\raisebox{7cm}{(b)}\\
\end{tabular}
\caption{Integration path for the channel Green operator $G^{ch}_{K^-n}$. See details in the text.}
\end{figure}

Similar considerations apply for the case of $G_3$ of eq.(\ref{gKp}):
\begin{equation*}
G_{3}(x_3,y_3;E)=\oint_c g_3(x_3;\varepsilon)g_0(y_3;E-\varepsilon)\ d\varepsilon\ .
\end{equation*}
When the original $\bar{K}N$ interaction couples the $\bar{K}N-\pi\Sigma-\pi\Lambda$ channels, as in one of our potential models, the optical potential
has two extra cuts and the above described procedure has to be applied to both of them, resulting in a somewhat more complicated integration path.

\subsection{Energy dependent potentials and the convolution integral}

The $\bar{K}N$ interactions enter the Faddeev equations in the form $G^{ch}_{K^-n}(x_2,y_2;E) v_{K^-n}(x_2)$ and $G_3(x_3,y_3;E)v_3(x_3)$. If the
potentials are energy-dependent either due to the optical potential construction or inherently (or both)
\begin{equation}
\label{endep}
v_{K^-n}(x_2)\Rightarrow v_{K^-n}(x_2;z)\qquad v_3(x_3)\Rightarrow  v_3(x_3;z)
\end{equation}
the convolution integral has to be modified. In eq.(\ref{endep}) the $z$ obviously refers to the corresponding two-body subsystem energy, which also occurs under the integration sign in eq.(\ref{conv}).
 Therefore the modified convolution integrals have the form
\begin{equation*}
 G^{ch}_{K^-n}(x_2,y_2;E) v_{K^-n}(x_2)\Rightarrow \oint_c g_{K^-n}(x_2;\varepsilon)v_{K^-n}(x_2;\varepsilon)g_c(y_2;E-\varepsilon)\ d\varepsilon\
\end{equation*}
and similarly for the other case. For separable interactions used in our calculation, this does not mean an extra difficulty, since for them $gv$ has a simpler form, than $g$ itself.

\section{Input and Results}
\label{Results.sect}
\subsection{$\bar{K}N$ interactions}
In our previous test calculation \cite{test} we used simple one-term separable interactions with complex coupling strengths to account for the absorption.
In the present calculation we used three different $\bar{K}N$ interactions, realistic in the sense, that they reproduce  all known $\bar{K}N$ experimental
data, including the recent SIDDHARTA value of the $1s$ level shift \cite{sid1} in kaonic hydrogen.

\vspace{.3cm}
\begin{tabular}{cl}
 $v_{\bar{K}N}({\rm SIDD1})\rightarrow$&$\bar{K}N-\pi\Sigma$ coupled channels, one-pole structure of the $\Lambda(1405)$\\
  $v_{\bar{K}N}({\rm SIDD2})\rightarrow$&$\bar{K}N-\pi\Sigma$ coupled channels, two-pole structure of the $\Lambda(1405)$\\
  $v_{\bar{K}N}({\rm Chiral})\rightarrow$&$\bar{K}N-\pi\Sigma-\pi\Lambda$ coupled channels, energy-dependent coupling \\
 &constants, channel couplings according to chiral perturbation\\&theory
\end{tabular}
\vspace{.3cm}

All interactions are separable with simple Yamaguchi form-factors, we have constructed them for our previous Coulombless Faddeev calculations for the $\bar{K}NN$ system, the SIDD1 and SIDD2 in \cite{prevb}, while the Chiral one in \cite{prevc}, where their detailed description can be found.

\subsection{$np$ interactions}
In order to have an idea about the effect of the deuteron structure on the level shift, we used two $np$ potentials:
\begin{itemize}
\item[(i)]{$v_{np}^s$ -- a simple one-term attractive separable potential, which reproduces the deuteron binding energy and size}
\item[(ii)]{$v_{np}^{a+r}$ -- a more realistic two-term attractive plus repulsive separable potential, reproducing the deuteron and the ${}^3S_1$ phase shifts up to 300 MeV}
\end{itemize}
\subsection{Results}

Our results for the $1s$  kaonic deuterium level shift
 $$\Delta E= E{\rm (3-body)}-E_0$$
are summarized in Table 1. The convergence of the method with increasing basis size is apparent, the accuracy of the
converged results is of the order of 1 eV. The different $\bar{K}N$ interactions, which are equally good in describing the two-body
data, give somewhat different level shifts,  the chiral value differs significantly from the two phenomenological ones. The deuteron
wave function (or the $np$ interaction) has also a certain, not too large, effect. Thus an available experimental value of $\Delta E$,
an expected and desired result of the SIDDHARTA 2 experiment \cite{sid2}, could contribute to our understanding of the $\bar{K}N$ interaction.

\begin{table*}
\begin{center}
\begin{tabular}{||c|c|c|c|c|c|c|c||}
\hline \noalign{\smallskip}
 \multicolumn{2}{||c|}{Basis} &
 \multicolumn{6}{c||
 }{$\Delta E$ in eV} \\
\hline \noalign{\smallskip}
$n_{max}$ in each &Total basis &\multicolumn{2}{c|}{$v_{\bar KN}$( SIDD1)}
&\multicolumn{2}{c|}{$v_{\bar KN}$(SIDD2)}&\multicolumn{2}{c||}{$v_{\bar KN}$( Chiral)} \\
\hhline{~~------}
channel&size $N_{tot}$&$v^{s}_{np}$&$v^{a+r}_{np}$&$v^{s}_{np}$&$v^{a+r}_{np}$&$v^{s}_{np}$&$v^{a+r}_{np}$\\
\noalign{\smallskip} \hline \noalign{\smallskip}
20 & $1764$  & $ 692-439 i$ & $714-452 i$  & $711-448 i$ & $ 728-448 i$ &  $ 762-461 i$ &  $ 766-460 i$  \\
24 & $2500$  & $ 699-442 i$ & $739-456 i$  & $738-451 i$ & $ 753-455 i$ &  $ 792-472 i$ &  $ 802-477 i$  \\
28 & $3364$  & $ 706-442 i$ & $753-459 i$  & $755-455 i$ & $ 769-461 i$ &  $ 809-480 i$ &  $ 823-490 i$  \\
32 & $4356$  & $ 711-442 i$ & $761-461 i$  & $765-458 i$ & $ 776-466 i$ &  $ 816-486 i$ &  $ 832-497 i$  \\
36 & $5476$  & $ 713-442 i$ & $764-463 i$  & $770-461 i$ & $ 780-468 i$ &  $ 819-489 i$ &  $ 835-500 i$  \\
40 & $6724$  & $ 715-442 i$ & $766-464 i$  & $774-461 i$ & $ 781-469 i$ &  $ 819-490 i$ &  $ 836-502 i$  \\
44 & $8100$  & $ 716-442 i$ & $767-464 i$  & $776-461 i$ & $ 782-469 i$ &  $ 820-491 i$ &  $ 835-502 i$  \\
\noalign{\smallskip} \hline
\end{tabular}
\end{center}
\caption{ Convergence of the kaonic deuterium $1s$ level shifts $\Delta E$ with increasing basis size $N_{tot}=4(n_{max}+1)^2$.
Results are shown for the three considered $\bar{K}N$ potentials $v_{\bar{K}N}({\rm SIDD1}), v_{\bar{K}N}({\rm SIDD2}),
v_{\bar{K}N}({\rm Chiral})$ and the two $np$ interactions $v_{np}^s$ and $v_{np}^{a+r}$.}
\label{res.tab}
\end{table*}

We also made a comparison of our converged results with some of the  approximations for  $\Delta E$, the results are shown in Table 2.
The "corrected Deser formula" \cite{Deser} connects $\Delta E$  with the strong scattering length $a_{\bar{K}d}$,
while in the ``best" two-body approximation \cite{prevb}, a strong $\bar{K}d$ optical  potential $V_{\bar{K}d}^{opt}$ is added to the Coulomb
interaction to calculate the shifted energy eigenvalue. For the numbers in Table 2. both $a_{\bar{K}d}$ and $V_{\bar{K}d}^{opt}$
were derived from the solution of Faddeev equations with the same strong potentials. It is evident, that the most commonly used
and often unduly trusted corrected Deser formula (in its most widely used form \cite{Boras}):
\begin{equation*}
\Delta E=-2 \alpha^3 \mu^2 a_{\bar{K}d}\ (1-2\ \alpha\  a_{\bar{K}d}\  \mu(\ln \alpha - 1)),
\end{equation*}
has little to do with the exact results, especially for the imaginary part of the level shift.
The ``best'' two-body approximation seems to give reasonable results, probably within the range of expected experimental
accuracy.

\begin{table*}
\begin{center}
\begin{tabular}{||c|c|c|c||}
\hline \noalign{\smallskip}
\multirow{2}{*}{$\bar{K}N$ potential}  & Corrected Deser  & $\bar Kd$ optical  &
\multirow{2}{*}{  3-body}    \\
&from $a_{Kd}$&potential& \\
\hline \noalign{\smallskip}
SIDD1 & $ 831 - 367 i $ & $ \ 785 - 509 i\ $ & $\  767-464 i\  $    \\
SIDD2 & $ 840 - 364 i $ & $ 797 - 512 i$  & $ 782-469 i$   \\
Chiral & $ 881 - 363 i $  & $ 828 - 527 i$  & $ 835-502 i$   \\

\noalign{\smallskip} \hline
\end{tabular}
\end{center}
\caption{ Comparison of calculation methods for $\Delta E$ (in $eV$)}
\end{table*}
\section{Conclusions}
\label{conclusions.sect}

\begin{itemize}
\item[(i)]{The present calculations, made with different, realistic  $\bar{K}N$ interactions suggest, that the level shift $\Delta E$ should be in the
range  $\Delta E\sim(800\pm30)-(480\pm20) i\ {\rm eV}$.}
\item[(ii)]{This is the first exact calculation of the level shift in a hadronic atom, which uses realistic, multichannel hadron-nucleon interaction and
 goes beyond the conventional two-body picture.}
\item[(iii)]{For the strangeness nuclear physics the main significance of the results is not as much in the obtained numbers, as in the first
 possibility to relate an important and hopefully measurable observable of the $\bar{K}NN$ system to the input $\bar{K}N$ interactions without
 relying upon uncontrollable approximations.}
 \item[(iv)]{The proposed method can serve as an important tool in fixing the yet  uncertain properties of the basic $\bar{K}N$ interactions.}
\end{itemize}
\vspace{5mm}

\noindent
{\bf Acknowledgments.}
The work was supported by the 
Hungarian OTKA grant 109462.

\end{document}